\documentclass[prd,twocolumn,amsmath,amssymb,superscriptaddress,floatfix,nofootinbib]{revtex4}

\usepackage{amsfonts}
\usepackage[dvipdfm,colorlinks=true,citecolor=blue,linkcolor=blue,urlcolor=blue]{hyperref}
\usepackage{color}
\usepackage{graphicx,amsfonts,multirow}
\usepackage{amsmath}
\usepackage{url}
\usepackage{ulem}

\usepackage{ulem}

\begin{document}

\title{The ratio $\mathcal{R}(D_s)$ for $B_s \to D_s \ell\nu_\ell$ by using the QCD light-cone sum rules within the framework of heavy quark effective field theory}

\author{Yi Zhang}
\email{yizhangphy@cqu.edu.cn}
\affiliation{Department of Physics, Chongqing University, Chongqing 401331, P.R. China}

\author{Tao Zhong}
\email{zhongtao1219@sina.com}
\affiliation{Department of Physics, Guizhou Minzu University, Guiyang 550025, P.R. China}

\author{Hai-Bing Fu}
\email{fuhb@cqu.edu.cn}
\affiliation{Department of Physics, Guizhou Minzu University, Guiyang 550025, P.R. China}

\author{Wei Cheng}
\email{chengwei@cqupt.edu.cn}
\affiliation{School of Science, Chongqing University of Posts and Telecommunications, Chongqing 400065, P.R. China}

\author{Long Zeng}
\email{zenglongz@outlook.com}
\affiliation{Department of Physics, Chongqing University, Chongqing 401331, P.R. China}

\author{Xing-Gang Wu}
\email{wuxg@cqu.edu.cn}
\affiliation{Department of Physics, Chongqing University, Chongqing 401331, P.R. China}
\affiliation{Chongqing Key Laboratory for Strongly Coupled Physics and Southwest Center for Theoretical Physics, Chongqing University, Chongqing 401331, P.R. China}

\date{\today}

\begin{abstract}

In the paper, we study the $B_s\to D_s$ transition form factors by using the light-cone sum rules within the framework of heavy quark effective field theory. We adopt a chiral current correlation function to do the calculation, the resultant transition form factors $f_+^{B_s\to D_s}(q^2)$ and $f_0^{B_s\to D_s}(q^2)$ are dominated by the contribution of $D_s$-meson leading-twist distribution amplitude, while the contributions from less certain $D_s$-meson twist-3 distribution amplitudes are greatly suppressed. At the largest recoil point, we obtain $f_{+,0}^{B_s \to D_s}(0)=0.533^{+0.112}_{-0.094}$. By further extrapolating the transition form factors into all the physically allowable $q^2$ region with the help of the $z$-series parametrization approach, we calculate the branching fractions $\mathcal{B}(B_s \to D_s \ell^\prime \nu_{\ell^\prime})$ with $(\ell^\prime= e,\mu)$ and $\mathcal{B}(B_s \to D_s \tau \nu_\tau)$, which gives $\mathcal{R}(D_s)=0.334\pm 0.017$.

\end{abstract}


\maketitle

\section {Introduction}\label{Section:I}

It is one of the most attractive research topics in the field of high energy physics to accurately test the standard model (SM) and to search new physics effects beyond the SM. The $B\to D^{(\ast)}$ semileptonic decay provides such an example. The ratio $\mathcal{R}(D^{(\ast)}) = \mathcal{B}(B\to D^{(\ast)}\tau\bar{\nu}_\tau) / \mathcal{B}(B\to D^{(\ast)} \ell^\prime \bar{\nu}_{\ell^\prime})$ with ($\ell^\prime=e,\mu$) has been measured by various groups, e.g. the BaBar Collaboration firstly reported $\mathcal{R}^{\rm exp.}(D) = 0.440\pm 0.058 \pm 0.042$ and $\mathcal{R}^{\rm exp.}(D^\ast) = 0.332 \pm 0.024 \pm 0.018$~\cite{BaBar:2012obs, BaBar:2013mob}, the BELLE Collaboration subsequently given $\mathcal{R}^{\rm exp.}(D) = 0.375\pm 0.064 \pm 0.026$ and $\mathcal{R}^{\rm exp.}(D^\ast) = 0.293 \pm 0.038 \pm 0.015$ in year 2015~\cite{Belle:2015qfa}, $\mathcal{R}^{\rm exp.}(D^\ast) = 0.270 \pm 0.035^{+0.028}_{-0.025}$ in year 2016~\cite{Belle:2016dyj,Belle:2017ilt}, and $\mathcal{R}^{\rm exp.}(D) = 0.307\pm 0.037 \pm 0.016$ and $\mathcal{R}^{\rm exp.}(D^\ast) = 0.283 \pm 0.018 \pm 0.014$ in year 2019~\cite{Belle:2019rba}, and the  LHCb Collaboration reported $\mathcal{R}^{\rm exp.}(D^\ast) = 0.336 \pm 0.027 \pm 0.030$ in year 2015~\cite{LHCb:2015gmp} and $\mathcal{R}^{\rm exp.}(D^\ast) = 0.283 \pm 0.018 \pm 0.014$ in year 2017~\cite{LHCb:2017smo, LHCb:2017rln}. The Heavy Flavor Average Group (HFLAG) gave the weighted average of those measurements, i.e. $\mathcal{R}^{\rm exp.}(D) = 0.339 \pm 0.026 \pm 0.014$ and $\mathcal{R}^{\rm exp.}(D^\ast) = 0.295 \pm 0.010 \pm 0.010$~\cite{HFLAV:2019otj}, where they also gave the averages of theoretical predictions $\mathcal{R}^{\rm th.}(D) = 0.298 \pm 0.003$ and  $\mathcal{R}^{\rm th.}(D^\ast) = 0.252 \pm 0.005$ from Refs.~\cite{Bigi:2016mdz, Bordone:2019vic, Gambino:2019sif}. Those theoretical values are consistent with other predictions calculated using various approaches, such as the heavy quark effective theory (HQET)~\cite{Fajfer:2012vx, Tanaka:2010se}, the lattice QCD (LQCD)~\cite{MILC:2015uhg, Na:2015kha, Aoki:2016frl}, the light-cone sum rules (LCSR)~\cite{Wang:2017jow, Zhong:2018exo}. Since the theoretical predictions are generally smaller than the measured ones, this difference may indicate new physics beyond the SM~\cite{Celis:2012dk, Celis:2013jha, Li:2016vvp, Li:2018lxi}.

The LHCb collaboration has measured the branching fraction $\mathcal{B}(B_s^0\to D_s^- \mu^+\nu_\mu)=(2.49 \pm 0.12 \pm 0.14 \pm 0.16) \times 10^{-2}$~\cite{LHCb:2020cyw} and gave the ratio of the branching fractions $\mathcal{B}(B_s^0\to D_s^- \mu^+\nu_\mu)$ and $\mathcal{B}(B^0\to D^- \mu^+\nu_\mu)$, i.e., $\mathcal{R} = 1.09 \pm 0.05 \pm 0.06 \pm 0.05$. This indicates $B_s\to D_s \ell\nu_\ell$ could behave closely to $B\to D \ell\nu_\ell$. Therefore, it is meaningful to make a detailed study on the similar ratio $\mathcal{R}(D_s)$.

At present, there is still no published data on the ratio $\mathcal{R}(D_s)$, while many theoretical studies on it have been done in Refs.\cite{Fan:2013kqa, Hu:2019bdf, Bhol:2014jta, Faustov:2012mt, Monahan:2017uby, Monahan:2018lzv, Dutta:2018jxz, McLean:2019qcx, Soni:2021fky}. As the key components of calculating the ratio $\mathcal{R}(D_s)$, the $B_s\to D_s$ transition form factors (TFFs) $f_+^{B_s\to D_s}(q^2)$ and $f_0^{B_s\to D_s}(q^2)$ have been studied under various approaches, e.g. the QCD sum rules (QCDSR)~\cite{Blasi:1993fi}, the constituent quark model (CQM)~\cite{Zhao:2006at}, the light-cone sum rules (LCSR)~\cite{Li:2009wq}, the Bethe-Salpeter equation (BSE)~\cite{Chen:2011ut}, and the lattice QCD (LQCD)~\cite{Monahan:2017uby, Monahan:2018lzv, Dutta:2018jxz, McLean:2019qcx}. Different approaches are applicable in various energy scale regions, for example, the LCSR is applicable in the largest low and intermediate $q^2$-region; and in the present paper, as the same as the previous treatment of $B\to\pi$ TFFs~\cite{Zhou:2019jny}, we will adopt the LCSR approach within the framework of heavy quark effective field theory (HQEFT)~\cite{Wu:1992zw, Wang:1999zd, Yan:1999kt, Wu:2000jq, Wang:2000sc, Wang:2000gs} to calculate $B_s\to D_s$ TFFs. The HQEFT separates the non-perturbative long-distance terms from the short-distance dynamics via a systematic way, and the long-distance terms can be decreased to a series over the non-perturbative wave functions or transition form factors. It has been pointed out that by choosing a proper chiral correlator, as will be adopted in this paper, one can suppress the uncertainties from the high-twist LCDAs and  achieve a more accurate LCSR prediction of the $B_s\to D_s$ TFFs.

The remaining parts of the paper are organized as follows. In Sec.~\ref{Section:II}, we present the calculation technologies for the two TFFs of the $B_s \to D_s \ell \bar{\nu}_\ell$ semileptonic decays by using the light-cone sum rules within the framework of HQEFT. In Sec.~\ref{Section:III}, we present our numerical results and discussions. Sec.~\ref{Section:IV} is reserved for a summary.

\section{Calculation technology}\label{Section:II}

\subsection{$B_s \to D_s$ Transition Matrix Element}

For the $B_s \to D_s l \bar{\nu}_l$ decays, the transition matrix element can be parameterized as follows:
\begin{eqnarray}
&&\langle D_s(p)|\bar{c} \gamma_\mu b |B_s(p+q)\rangle \nonumber\\
&=& 2 f_+^{B_s \to D_s}(q^2) p_\mu + \left[f_+^{B_s \to D_s}(q^2)+ f_-^{B_s \to D_s}(q^2) \right] q_\mu   \label{Eq:matrix1}
\end{eqnarray}
and
\begin{eqnarray}
f_0^{B_s \to D_s}(q^2) = f_+^{B_s \to D_s}(q^2)+ \frac{q^2}{m_{B_s}^2-m_{D_s}^2} f_-^{B_s \to D_s}(q^2),  \label{Eq:f0}
\end{eqnarray}
where $p$ is the momentum of the $D_s$-meson and $(p+q)$ is the momentum of $B_s$-meson. At the maximum recoil point, we have $f_{+}^{B_s \to D_s}(0) = f_{0}^{B_s \to D_s}(0)$. The transition matrix element can be expanded as $1/m_b$-power series within the framework of HQEFT. Based on the heavy quark symmetry, the transition matrix element of heavy quark in the effective theory is parameterized as~\cite{Wang:1999zd, Wang:2000sc, Wang:2000gs}:
\begin{eqnarray}
\langle D_s(p)| \bar{c} \gamma_\mu b |B_s(p+q)\rangle
&=& \frac{\sqrt{m_{B_s}}}{\sqrt{\bar\Lambda_{B_s}}}\langle D_s(p)|\bar u \gamma _\mu b_v^+ |{B_s}_v \rangle \nonumber\\
&=& - \frac{\sqrt{m_{B_s}}}{\sqrt{\bar\Lambda_{B_s}}}{\rm Tr}[D_s(v,p){\gamma _\mu }{\mathcal{M}}_v], \label{Eq:Bpi_HQEFT}
\end{eqnarray}
where
\begin{eqnarray}
\bar \Lambda_{B_s} &=& m_{B_s} - m_b, \nonumber\\
D_s (v,p) &=& {\gamma ^5}[A(v \cdot p)+ \not\!\hat{p} B(v \cdot p)], \\
{\mathcal{M}}_v &=& -\sqrt {\bar \Lambda }(1 + \not\! v)\gamma^5/2, \nonumber
\end{eqnarray}
where $b_v^+$ is the effective $b$-quark field and $v$ is the $B_s$-meson velocity, $\hat{p}^{\mu}=p^{\mu}/(v\cdot p)$. $A(v \cdot p)$ and $B(v \cdot p)$ are leading-order heavy flavor-spin independent coefficient functions. $\bar \Lambda =\mathop {\lim }\limits_{m_b \to \infty } {\bar \Lambda _{B_s}}$, which is the heavy flavor independent binding energy that reflects the effects of light degrees of freedom in the heavy hadron. Using those formulas, we obtain the $B_s \to D_s\ell\bar\nu_\ell$ TFFs $f_\pm(q^2)$, which are
\begin{eqnarray}
f_\pm^{B_s \to D_s}(q^2) = \frac{\sqrt {\bar\Lambda}}{\sqrt{m_{B_s}\bar\Lambda_{B_s}}}\left[A(y)\pm \frac{m_{B_s}}{y}B(y)\right]+ \cdots, \label{Eq:fpm}
\end{eqnarray}
with
\begin{equation}
y= v \cdot p=(m^2_{B_s} + m^2_{D_s} - q^2)/(2 m_{B_s}),
\end{equation}
where ``$\cdots$" denotes the higher-order ${\cal O}(1/{m_b})$ contributions that will not be taken into consideration here.

\subsection{Light-Cone Sum Rule For $f_{+,0}^{B_s \to D_s}(q^2)$}

To derive the sum rules of the two leading order heavy flavor-spin independent coefficient functions $A(y)$ and $B(y)$, we construct the following correlator:
\begin{eqnarray}
&& F_\mu (p,q) = i\int d^4x e^{iq \cdot x}\langle D_s (p)|T\{ j_n(x),j^\dag_0(0)\} |0\rangle, \label{Eq:correlator}
\end{eqnarray}
where the currents
\begin{eqnarray}
j_n(x) &=& \bar c(x)\gamma _\mu (1 + \gamma_5)b(x)  \\
j^{\dag}_0(0)&=& \bar b(0) i (1 + \gamma _5)s(0)
\end{eqnarray}

Following the standard procedure of LCSR approach, we first deal with the hadronic representation for the correlation function. One can insert a complete series of the intermediate hadronic states in the correlator~\eqref{Eq:correlator} in the physical $q^2$-region and isolate the pole term of the lowest pseudoscalar state from the hadronic representation. Then the correlator $F_\mu(p,q)$ becomes:
\begin{eqnarray}
&& F_\mu^{\rm Had.} (p,q)= \frac{ \langle D_s(p)|\bar c \gamma_\mu b |B_s \rangle \langle B_s | \bar b i \gamma_5 s | 0 \rangle }{m_{B_s}^2 - (p + q)^2} \cr
&&\quad + \sum \limits_{B_s^H} \frac{\langle D_s (p)|\bar c \gamma _\mu (1 + \gamma_5) b | B_s^H \rangle \langle B_s^H|\bar b i (1 + \gamma_5) s|0\rangle }{m_{B_s^H}^2 - (p + q)^2}.\nonumber\\ \label{Eq:hadronic}
\end{eqnarray}

In the effective theory of heavy quark, the hadronic representation~\eqref{Eq:hadronic} can be expanded in powers of  $1/m_b$. Taking the transition matrix element~\eqref{Eq:Bpi_HQEFT} into consideration and neglecting the contributions from higher $1/m_b$ order, we can further write the hadronic representation as:
\begin{eqnarray}
F_\mu^{\rm Had.} (p,q) &=& 2 F \frac{A(y)v^\mu  + B(y) \hat p ^\mu }{2 \bar\Lambda_{B_s} - 2v \cdot k} \cr
&& + \int_{s_0}^\infty  ds \frac{\rho (y,s)}{s - 2v \cdot k} + {\rm Subtractions}, \label{Eq:Fpq_HQEFT}
\end{eqnarray}
with the matrix element~\cite{Korner:2002ba}
\begin{eqnarray}
\langle B_s |\bar b^+_v i \gamma_5 d | 0 \rangle  = \frac{i}{2} F {\rm Tr}[\gamma_5 \mathcal{M}_v],
\end{eqnarray}
where $F$ is the leading-order decay constant of the $B_s$-meson~\cite{Wang:2001mi, Wang:2002zba}. $k$ is the residual momentum of the heavy hadronic. Using the ansatz of the quark-hadron duality the spectral density $\rho (y,s)$ can be obtained~\cite{Shifman:1978bx, Shifman:1978by}.

On the other hand, we apply the operator product expansion (OPE) to the correlator in the deep Euclidean region. The correlator (\ref{Eq:correlator}) can be explicitly written as
\begin{eqnarray}
F_\mu (p,q) &=& i\int d^4x e^{i(q-m_b) \cdot x}\langle D_s (p)|T\{ \bar c(x)\gamma _\mu (1 + \gamma_5)b^+_{v}(x), \nonumber\\
&&\bar b^+_{v}(0) i (1 + \gamma _5)s(0)\} |0\rangle. \label{Eq:correlator}
\end{eqnarray}
Using the $B$-meson heavy-quark propagator $S(x,v)=(1+\slash\!\!\! v )\times\int_0^\infty dt \delta(x-vt)/2$~\cite{Wang:2001mi}, the correlator can be expanded as a complex power series over the $D_s$-meson LCDAs. Due to the chiral suppressions, it is noted that the main contribution to the correlator comes from the leading-twist LCDA, and the contributions from all the twist-3 LCDAs are exactly zero.

Through the dispersion relation, the OPE in deep Euclidean region and the hadron expression in physical region can be matched. And by further applying the Borel transformation to suppress the contributions from power-suppressed terems, the LCSRs for  the coefficient functions $A(y)$ and $B(y)$ are
\begin{eqnarray}
A(y) &=& -\frac{f_{D_s}}{2F}\int_0^{s_0^{B_s}} ds e^{(2 \bar \Lambda _{B_s} - s)/T} \frac{1}{y^2} \frac{\partial}{\partial u} g_2(u) \bigg|_{u = 1 - \frac{s}{2y}},
\label{Eq:Ay}\\
B(y) &=& -\frac{f_{D_s}}{2F}\int_0^{s_0^{B_s}} ds e^{(2 \bar \Lambda _{B_s} - s)/T} \bigg[- \phi_{2;D_s}(u) \nonumber\\
&& + \bigg(\frac{1}{y}\frac{\partial }{\partial u}\bigg)^2 g_1(u) - \frac{1}{y^2}\frac{\partial}{\partial u}g_2(u)\bigg]\bigg|_{u = 1 - \frac{s}{2y}}, \label{Eq:By}
\end{eqnarray}
where $T$ is the Borel parameter and $s_0^{B_s}$ is the continuum threshold, $g_1$ and $g_2$ are twist-four LCDAs. Since the contributions from the twist-4 LCDAs are only several percent, so we shall directly adopt the light pseudo-scale ones to do the calculations, whose explicit forms can be found in Refs.\cite{Ball:1998je}. Substituting them into Eqs.(\ref{Eq:f0},\ref{Eq:fpm}), we obtain
\begin{eqnarray}
f_+^{B_s \to D_s}(q^2) &=&  - \frac{f_{D_s}\sqrt{\bar\Lambda}}{2F\sqrt{m_{B_s}\bar\Lambda_{B_s}}}  \int_0^{s_0^{B_s}} ds e^{(2 \bar \Lambda_{B_s} - s)/T} \nonumber\\
&\times&\bigg\{\frac{1}{y^2}\frac{\partial }{\partial u}g_2(u) + \frac{m_{B_s}}{y} \bigg[ - \phi_{2;D_s}(u) \nonumber\\
&+& \bigg(\frac{1}{y} \frac{\partial}{\partial u}\bigg)^2 g_1 (u) - \frac{1}{y^2}\frac{\partial }{\partial u}g_2(u)\bigg]\bigg\}\bigg|_{u = 1 - \frac{s}{2y}},  \label{Eq:f+}\\
f_0^{B_s \to D_s}(q^2) &=&  - \frac{f_{D_s}\sqrt{\bar\Lambda}}{2F\sqrt{m_{B_s}\bar\Lambda_{B_s}}}  \int_0^{s_0^{B_s}} ds e^{(2 \bar \Lambda_{B_s} - s)/T} \nonumber\\
&\times&\bigg\{\bigg(1 + \frac{q^2}{m_{B_s}^2 - m_{D_s}^2}\bigg)\frac{1}{y^2}\frac{\partial }{\partial u}g_2(u) \nonumber\\
&+& \bigg(1 - \frac{q^2}{m_{B_s}^2-m_{D_s}^2}\bigg)\frac{m_{B_s}}{y} \bigg[ - \phi_{2;D_s}(u) \nonumber\\
&+& \bigg(\frac{1}{y} \frac{\partial}{\partial u}\bigg)^2 g_1 (u) - \frac{1}{y^2}\frac{\partial }{\partial u}g_2(u)\bigg]\bigg\}\bigg|_{u = 1 - \frac{s}{2y}}.\label{Eq:f01}
\end{eqnarray}
The Borel parameter $T$ and  the continuum threshold $s_0^{B_s}$ shall be fixed such that the resulting TFFs do not depend too much on the precise values of those parameters. In addition, the continuum contribution, which is the part of the dispersive integral from $s_0^{B_s}$ to $\infty$ that is subtracted from both sides of the equation, should not be too large.

\section{Numerical analysis}\label{Section:III}

\subsection{Input parameters}

To determine the TFFs $f_{+,0}^{B_s \to D_s}(q^2)$ of the exclusive process $B_s \to D_s \ell \bar\nu_\ell$, we take~\cite{Zyla:2020zbs,Wang:2000sc}
\begin{eqnarray}
m_{B_s}&=&5.367 \pm 0.00014 {\rm GeV},   \nonumber\\
m_{D_s}&=&1.968 \pm 0.00007 {\rm GeV},   \nonumber\\
f_{D_s}&=&0.256  \pm 0.0042 {\rm GeV}, \nonumber\\
F&=&0.30 \pm 0.04 {\rm GeV^{3/2}}. \nonumber
\end{eqnarray}

\begin{table}[htb]
\caption{The $D_s$-meson leading-twist LCDA parameters at the scale $\mu =3{\rm GeV}$. The values in the second row are for the central values, and the values in third and fourth rows are uncertainties.}
\begin{tabular}{cccccc}
\hline
~$A_{D_s}({\rm GeV}^{-1})$~ & ~$B^{D_s}_1$~ & ~$B^{D_s}_2$~& ~$B^{D_s}_3$~ & ~$B^{D_s}_4$~ & ~$\beta_{D_s}({\rm GeV})$~ \\
\hline
~$1.246$~ & ~$-0.214$~ & ~$-0.167$~& ~$0.055$~ & ~$0.005$~ & ~$5.521$~ \\
~$11.001$~ & ~$-0.165$~ & ~$0.014$~& ~$-0.004$~ & ~$0.003$~ & ~$1.046$~ \\
~$1.184$~ & ~$-0.189$~ & ~$-0.163$~& ~$0.047$~ & ~$0.008$~ & ~$6.970$~ \\
\hline
\end{tabular}
\label{DAparameter}
\end{table}

\begin{figure}[htb]
\begin{center}
\includegraphics[width=0.45\textwidth]{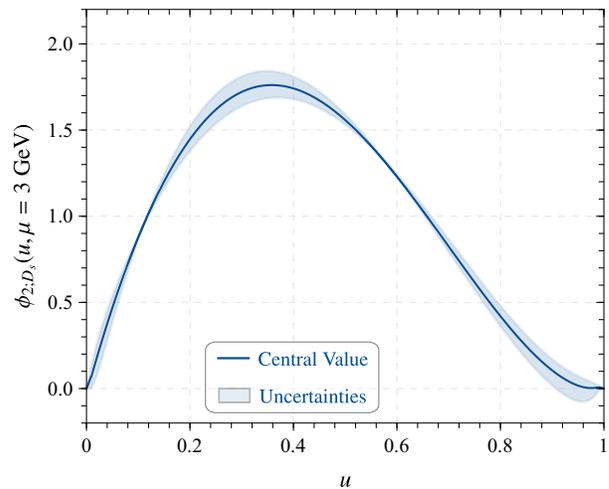}
\end{center}
\caption{The behavior of LCDA $\phi_{2;Ds}(u,\mu=3{\rm GeV})$, the shaded area indicates LCDA's uncertainties}. \label{DA}
\end{figure}

For the $D_s$-meson leading-twist LCDA $\phi_{2;D_s}(u,\mu)$ in the LCSRs of TFFs \eqref{Eq:f+} and \eqref{Eq:f01}, we adopt the light cone harmonic oscillator (LCHO) model suggested in Ref.\cite{Zhang:2021wnv}, which has a better end-point behavior and takes the following form:
\begin{eqnarray}
\phi_{2;D_s}(u,\mu) &=& \frac{\sqrt{6} A_{D_s} \beta_{D_s}^2}{\pi^2 f_{D_s}} u\bar{u} \varphi_{2;D_s}(u) \nonumber\\
&\times& \exp \left[ - \frac{\hat{m}_c^2u + \hat{m}_s^2\bar{u}}{8\beta_{D_s}^2 u\bar{u}} \right] \nonumber\\
&\times& \left\{ 1 - \exp \left[ -\frac{\mu^2}{8\beta_{D_s}^2 u\bar{u}} \right] \right\},
\label{phi}
\end{eqnarray}
where $\bar{u} = 1 - u$ and $\varphi_{2;D_s}(u) = 1 + \sum^{4}_{n=1} B_n^{D_s} \times C_n^{3/2}(\xi)$ with $\xi = u-\bar{u}$. $\hat{m}_{c}$ and $\hat{m}_{s}$ are $c$- and $s$-constituent quark masses, whose values are taken as $\hat{m}_{c}\simeq 1.5~{\rm GeV}$ and $\hat{m}_{s}\simeq 0.5~{\rm GeV}$. $A_{D_s}$. $\beta_{D_s}$, $B_1^{D_s}$, $B_2^{D_s}$, $B_3^{D_s}$ and $B_4^{D_s}$ are model parameters, whose initial values at the scale $\mu=2 {\rm GeV}$ have been given in Ref.~\cite{Zhang:2021wnv}. For the present process, the typical factorization scale $\mu$ = $(m^2_{B_s}-m_b^2)^{1/2}=3 {\rm GeV}$. The input parameters at the scale $\mu=3 {\rm GeV}$ can be achieved by using the conventional one-loop evolution equation~\cite{Lepage:1980fj}, and these values are given in Table~\ref{DAparameter}. Figure~\ref{DA} shows the behavior of the $D_s$-meson leading-twist LCDA $\phi_{2;D_s}(u,\mu=3{\rm GeV})$ with the typical values exhibited in Table~\ref{DAparameter}, where the solid line is the central value and the shaded band shows its uncertainty given in Table~\ref{DAparameter}.

\subsection{The $B_s \to D_s$ TFFs}

\begin{figure}[htb]
\begin{center}
\includegraphics[width=0.45\textwidth]{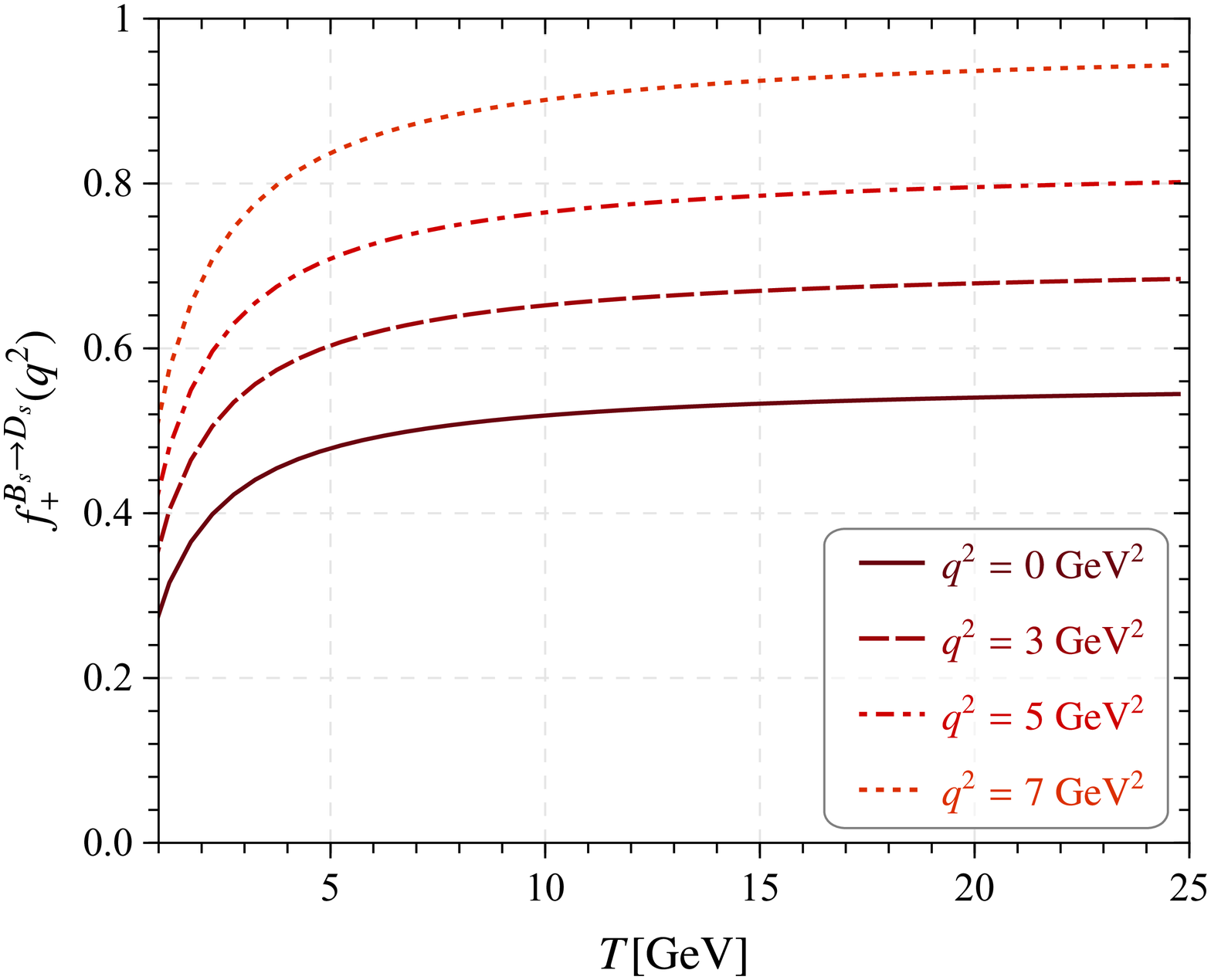}
\includegraphics[width=0.45\textwidth]{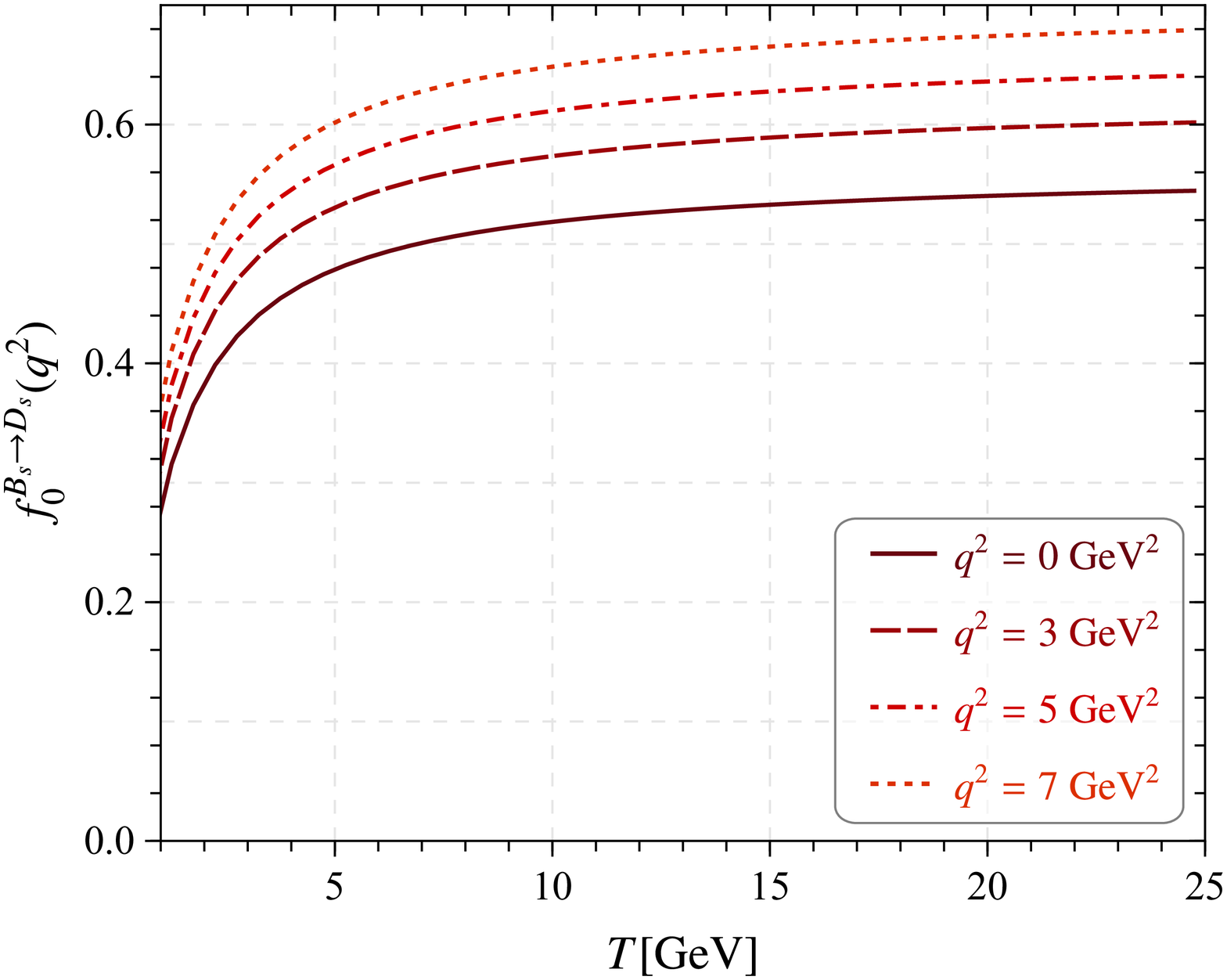}
\end{center}
\caption{The TFFs $f_+^{B_s\to D_s}(q^2)$ and $f_0^{B_s\to D_s}(q^2)$ versus the Borel parameter $T$, where several typical $q^2$ are adopted and all input parameters have been set as their central values.} \label{Aqr}
\end{figure}

Next, we calculate the $B_s\to D_s$ TFFs $f^{B_s\to D_s}_+(q^2)$ and $f^{B_s\to D_s}_0(q^2)$ by using the LCSRs \eqref{Eq:f+} and \eqref{Eq:f01} in the $q^2$-region when the LCSR approach is applicable, i.e., $0<q^2<7{\rm GeV^2}$. For the purpose, we first determine the continuum threshold $s_0^{B_s}$ and the Borel parameter $T$. Within the framework of HQEFT, the continuum threshold $s_0^{B_s} \equiv 2\bar{\Lambda}_{B_s^*} = 2(m_{B_s^*} - m_b)$ with $B_s^*$ being the $B_s$-meson first excited state~\cite{Wang:2001mi}, and we take $s_0^{B_s}=3.85 \pm 0.15{\rm GeV}$. To determine the Borel window, as suggested in Refs.\cite{Wang:2001mi, Wang:2002zba}, we require the TFFs $f^{B_s\to D_s}_{+,0}(q^2)$ to be as stable as possible within corresponding Borel windows. Figure \ref{Aqr} shows the TFFs $f^{B_s\to D_s}_{+,0}(q^2)$ versus the Borel parameter $T$ at several typical squared momenta transfer, in which the solid, the dashed, the dot-dashed and the dotted lines are for $q^2 = 0, 3, 5, 7{\rm GeV}^2$, respectively. One can find that, the TFFs $f^{B_s\to D_s}_{+,0}(q^2)$ are stable for a large $T$, e.g. the uncertainty caused by $T$ is less than $5\%$ when $T \geq 10{\rm GeV}$ for all those $q^2$ values. Therefore, we take the Borel window as $10 {\rm GeV}<T<20 {\rm GeV}$.

\begin{table}[htb]
\caption{Theoretical predictions of the TFF $f_{+,0}^{B_s \to D_s}(q^2)$ at the maximum recoil point under various approaches.}
\begin{tabular}{l l}
\hline
~~Methods~~~~~~~~~~~~~~~~~~~~~~~~~~~~~~~~~~~~~~~~~~ & ~$f_{+,0}^{B_s \to D_s}(0)$~ \\
\hline
~This work (HQEFT)~ & ~$0.533^{+0.112}_{-0.094}$~ \\
~pQCD~\cite{Fan:2013kqa}~ & ~$0.55^{+0.15}_{-0.12}$~ \\
~pQCD+LQCD~\cite{Hu:2019bdf}~ & ~$0.52 \pm 0.10$~ \\
~RQM~\cite{Bhol:2014jta}~ & ~$0.74 \pm 0.02$~ \\
~LQCD~\cite{Monahan:2017uby}~ & ~$0.656\pm0.031$~ \\
~LQCD~\cite{Monahan:2018lzv}~ & ~$0.661\pm 0.042$~ \\
~QCDSR~\cite{Blasi:1993fi}~ & ~$0.7 \pm 0.1$~ \\
~QCDSR~\cite{Azizi:2008tt}~ & ~$0.24$~ \\
~LCSR~\cite{Li:2009wq}~ & ~$0.43^{+0.09}_{-0.08}$~ \\
~BSE~\cite{Chen:2011ut}~ & ~$0.57^{+0.02}_{-0.03}$~ \\
\hline
\end{tabular}
\label{TFF0}
\end{table}

At the maximum recoil point $q^2=0$, we have
\begin{eqnarray}
f^{B_s \to D_s}_{+,0}(0) &=& 0.533^{+0.082}_{-0.063}|_{\phi_{2;D_s}}\ ^{+0.007}_{-0.014}|_{T}\ ^{+0.065}_{-0.063}|_{s_0^{B_s}}\nonumber\\
 &&^{+0.027}_{-0.037}|_{F}\ ^{+0.004}_{-0.004}|_{m_b}.
 \label{Eq:f+00}
\end{eqnarray}
There are also errors caused by the uncertainties of $m_{B_s}$ and $m_{D_s}$, which are negligibly small. It is fond that the uncertainties of the $D_s$-meson leading-twist LCDA $\phi_{2;D_s}$ and the continuum threshold $s_0^{B_s}$ are main errors of $f^{B_s \to D_s}_{+,0}(0)$. By  adding all those errors in quadrature, we obtain $f^{B_s \to D_s}_{+,0}(0) = 0.533^{+0.160}_{-0.128}$. We present the theoretical predictions of $f_{+,0}^{B_s \to D_s}(q^2)$ at the maximum recoil point $q^2=0$ in Table~\ref{TFF0}, where the predictions under the pQCD approach~\cite{Fan:2013kqa}, the pQCD+LQCD approach~\cite{Hu:2019bdf}, the LQCD approach~\cite{Monahan:2017uby, Monahan:2018lzv}, the QCD SR approach~\cite{Blasi:1993fi, Azizi:2008tt}, the LCSR approach~\cite{Li:2009wq}, the BSE approach~\cite{Chen:2011ut} and the RQM approach~\cite{Bhol:2014jta} are also presented. Our present prediction of $f^{B_s \to D_s}_{+,0}(0)$ is in good agreement with the values calculated with the pQCD prediction~\cite{Fan:2013kqa}, the pQCD+LQCD prediction~\cite{Hu:2019bdf} and the BSE prediction~\cite{Chen:2011ut}.

\begin{figure}[htb]
\begin{center}
\includegraphics[width=0.45\textwidth]{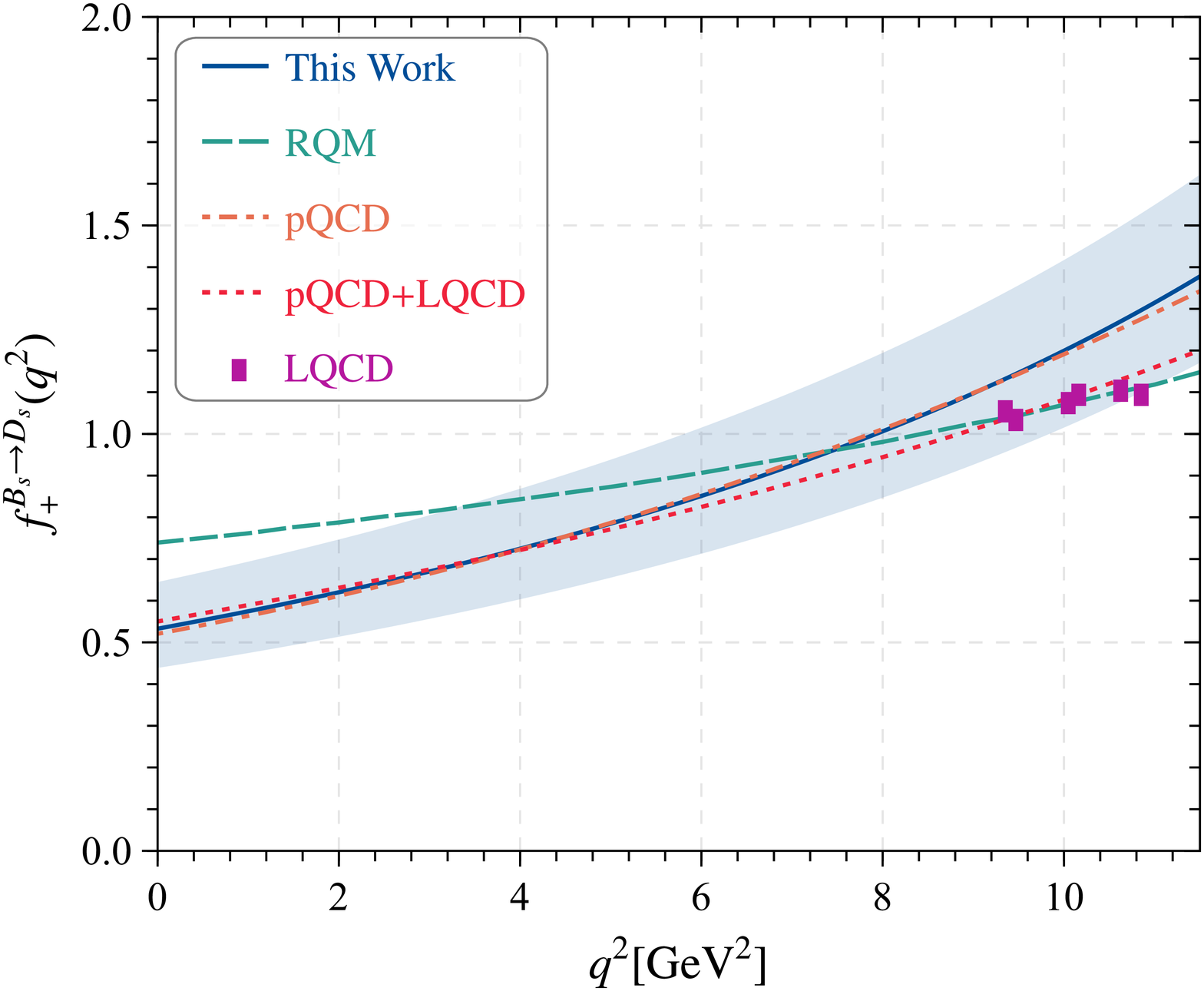}
\includegraphics[width=0.45\textwidth]{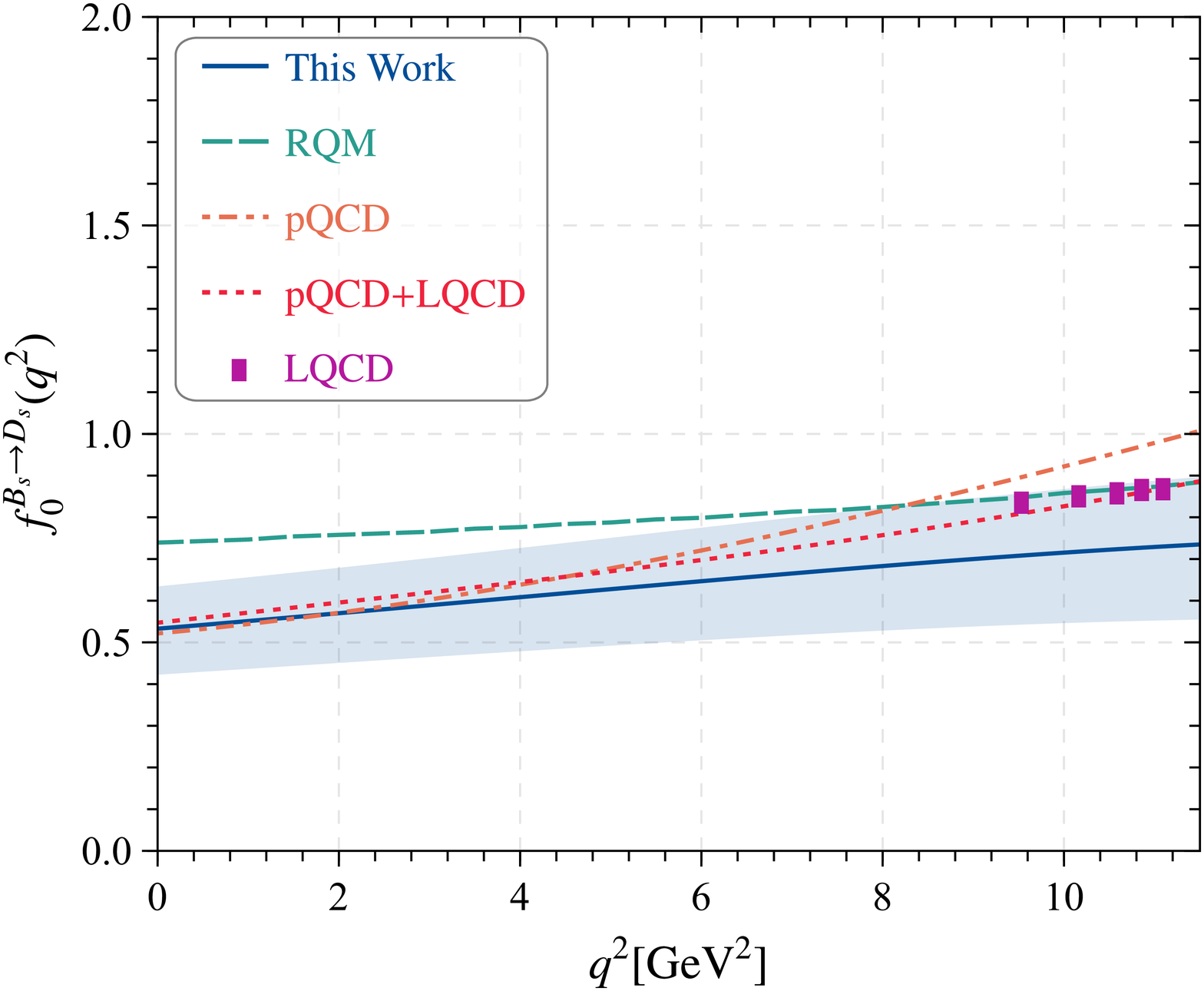}
\end{center}
\caption{The extrapolated TFFs $f_{+,0}^{B_s \to D_s}(q^2)$ versus $q^2$. The solid line are central values and the shaded bands are corresponding uncertainties. As a comparison, the predictions under the pQCD+LQCD approach~\cite{Hu:2019bdf}, the pQCD approach~\cite{Hu:2019bdf}, and the RQM approach~\cite{Faustov:2012mt} and the LQCD approach~\cite{Monahan:2017uby} are also presented.}
\label{fp}
\end{figure}

As mentioned above, the LCSRs \eqref{Eq:f+} and \eqref{Eq:f01} for TFFs $f^{B_s\to D_s}_{+,0}(q^2)$ are only reliable in low and intermediate regions, i.e., $0<q^2<7{\rm GeV^2}$. To estimate the total decay width of the semi-leptonic decay $B_s \to D_s \ell\bar \nu_\ell$, we extrapolate the TFFs to the whole physically allowable $q^2$-region, $0<q^2<(m_{B_s}-m_{D_s})^2=11.50{\rm GeV^2}$, via the $z$-series parametrization~\cite{Khodjamirian:2011ub, Bourrely:2008za}:
\begin{eqnarray}
f_+^{B_s \to D_s} (q^2) &=& \frac{f_+^{B_s \to D_s} (0)}{1 - q^2/m_{B_s^*}^2}\bigg\{ 1 + \sum_{k = 1}^{N - 1} b_k \bigg[z(q^2)^k - z(0)^k\nonumber\\
 &&- ( - 1)^{N - k}\frac{k}{N}(z(q^2)^N - z(0)^N)\bigg]\bigg\}, \label{Eq:f+EX}\\
 f_0^{B_s \to D_s}(q^2) &=& f_0^{B_s \to D_s}(0)\bigg\{ 1 + \sum_{k = 1}^{N - 1} b_k (z(q^2)^k - z(0)^k)\bigg\}. \nonumber \\ \label{Eq:f0EX}
\end{eqnarray}
where
\begin{eqnarray}
z(q^2) &=& \frac{\sqrt {t_ +  - q^2}  - \sqrt {t_ +  - t_0} }{\sqrt {t_ + - q^2}  + \sqrt {t_ + - t_0} }, \nonumber\\
t_0 &=& t_ + (1 - \sqrt {1-t_+/t_-} ), \nonumber\\
t_ \pm &=& (m_{B_s} \pm m_{D_s})^2. \nonumber
\end{eqnarray}
Then, by fitting the values of the TFFs in low and intermediate regions calculated via the LCSRs \eqref{Eq:f+} and \eqref{Eq:f01}, the coefficients $b_1$, $b_2$ and $b_3$ in extrapolation formula \eqref{Eq:f+EX} and \eqref{Eq:f0EX} can be determined, and which have been exhibited in Table~\ref{tbw}. The quality-of-fit is defined as:
\begin{equation}
\Delta  = \frac{\sum_t {|f^{i} - f^{\rm fit}|}}{\sum_t {|f^{i}|} } \times 100\%,t \in \bigg\{ 0,\frac{1}{2},...,\frac{23}{2},12\bigg\} ~\rm{GeV}^2.
\end{equation}
The coefficients $b_i$ are determined such that the quality-of-fit $(\Delta)$ is no more than $1\%$. The $\Delta$ values for the central, the upper and lower TFFs are shown in Table~\ref{tbw}. These quality-of-fits are much smaller than $1\%$, indicating that our present extrapolations are of high accuracy. We present the extrapolated TFFs $f^{B_s\to D_s}_{+,0}(q^2)$ in Figure~\ref{fp}, where the shaded hands are theoretical uncertainties from all the mentioned error sources. For comparison, we present the results of the pQCD+LQCD approach~\cite{Hu:2019bdf}, the pQCD approach~\cite{Hu:2019bdf}, the RQM approach~\cite{Faustov:2012mt} and the LQCD approach~\cite{Monahan:2017uby}.

\begin{table}[htb]
\caption{The fitted parameters and the quality-of-fit for the extrapolated TFFs $f^{B_s\to D_s}_{+,0}(q^2)$.}
\begin{tabular}{ l  c  c  c  c}
\hline
~$f_0^{B_s \to D_s}(q^2)$~~~~~~~~~~~ & ~$b_1$~ & ~$b_2$~ & ~$b_3$~ & ~$(\Delta)$~ \\
\hline
~$0.533$~ & ~$-2.378$~ & ~$-19.414$~& ~$196.397$~& ~$0.006\%$~ \\
~$0.533^{+0.112}$~ & ~$-3.586$~ & ~$-15.843$~& ~$236.929$~& ~$0.006\%$~ \\
~$0.533_{-0.094}$~ & ~$-4.646$~ & ~$-12.450$~& ~$281.244$~& ~$0.006\%$~ \\
\hline
~$f_+^{B_s \to D_s}(q^2)$~ & ~$b_1$~ & ~$b_2$~ & ~$b_3$~ & ~$(\Delta)$~ \\
\hline
~$0.533$~ & ~$-4.389$~ & ~$-0.641$~& ~$123.375$~& ~$0.006\%$~ \\
~$0.533^{+0.112}$~ & ~$-4.888$~ & ~$-1.687$~& ~$165.880$~& ~$0.008\%$~ \\
~$0.533_{-0.094}$~ & ~$-5.459$~ & ~$-4.242$~& ~$209.787$~& ~$0.006\%$~ \\
\hline
\end{tabular}
\label{tbw}
\end{table}

\subsection{The $B_s \to D_s \ell\bar \nu_\ell $ branching fractions and the ratio $\mathcal{R}(D_s)$}

The branching fraction of the semileptonic decay $B_s\to D_s \ell\bar{\nu}_\ell$ is defined as
\begin{equation}
\mathcal{B}(B_s\to D_s \ell\bar{\nu}_\ell) = \tau_{B_s} \times \int^{(m_{B_s} - m_{D_s})^2}_{0} \!\!\!dq^2 \frac{d\Gamma(B_s\to D_s \ell\bar{\nu}_\ell)}{dq^2}, \label{Eq:BF}
\end{equation}
where $q_{\rm max}^2 = (m_{B_s} - m_{D_s})^2$ and $\tau_{B_s}$ is the $B_s$-meson lifetime. Here the differential decay widths is
\begin{eqnarray}
\frac{d\Gamma(B_s \to D_s \ell\bar \nu_\ell )}{dq^2}&=&\frac{G_F^2|V_{cb}|^2}{192 \pi^3 m_{B_s}^3} \left(1- \frac{m_\ell^2}{q^2}\right)^2 \left[\left( 1 + \frac{m_\ell^2}{2 q^2} \right)\right.\nonumber\\
&&\times \lambda^\frac{3}{2} (q^2) |f_+^{B_s \to D_s}(q^2)|^2 \nonumber\\
&& + \frac{3m^2_\ell}{2q^2}(m^2_{B_s}-m^2_{D_s})^2  \nonumber\\
&& \times\left.   \lambda^\frac{1}{2} (q^2) |f_0^{B_s \to D_s}(q^2)|^2 \right], \label{Eq:DDF1}
\end{eqnarray}
where $\lambda(q^2) = (m_{B_s}^2 + m_{D_s}^2 - q^2)^2 - 4m_{B_s}^2 m_{D_s}^2$, which is the phase-space factor. $|V_{cb}|$, $G_F$ and $m_\ell$ are CKM matrix element, the Fermi-coupling constant and the lepton mass, respectively, and we take~\cite{Zyla:2020zbs}: $\tau_{B_s}=(1.510\pm{0.004}) \times 10^{-12}s$, $|V_{cb}|=(40.5 \pm 1.5) \times 10^{-3}$, $G_F=1.1663787(6)\times 10^{-5}{ \rm GeV}^{-2}$ and $\tau$-lepton mass $m_\tau=1.776 \pm 0.00012{ \rm GeV}$. For lepton $\ell^{\prime}=e$ or $\mu$, its mass is negligible, and then the above differential decay width can be simplified as
\begin{equation}
\frac{d\Gamma(B_s \to D_s \ell^{\prime}\bar \nu_{\ell^{\prime}} )}{dq^2}= \frac{G_F^2
  |V_{cb}|^2}{ 192 \pi^3 m_{B_s}^3}\,\lambda^{3/2}(q^2) |f_+^{B_s \to D_s}(q^2)|^2, \label{Eq:DDF2}
\end{equation}
where $f_0^{B_s \to D_s}(q^2)$ has zero contribution due to chiral suppression.

\begin{figure}[htb]
\begin{center}
\includegraphics[width=0.45\textwidth]{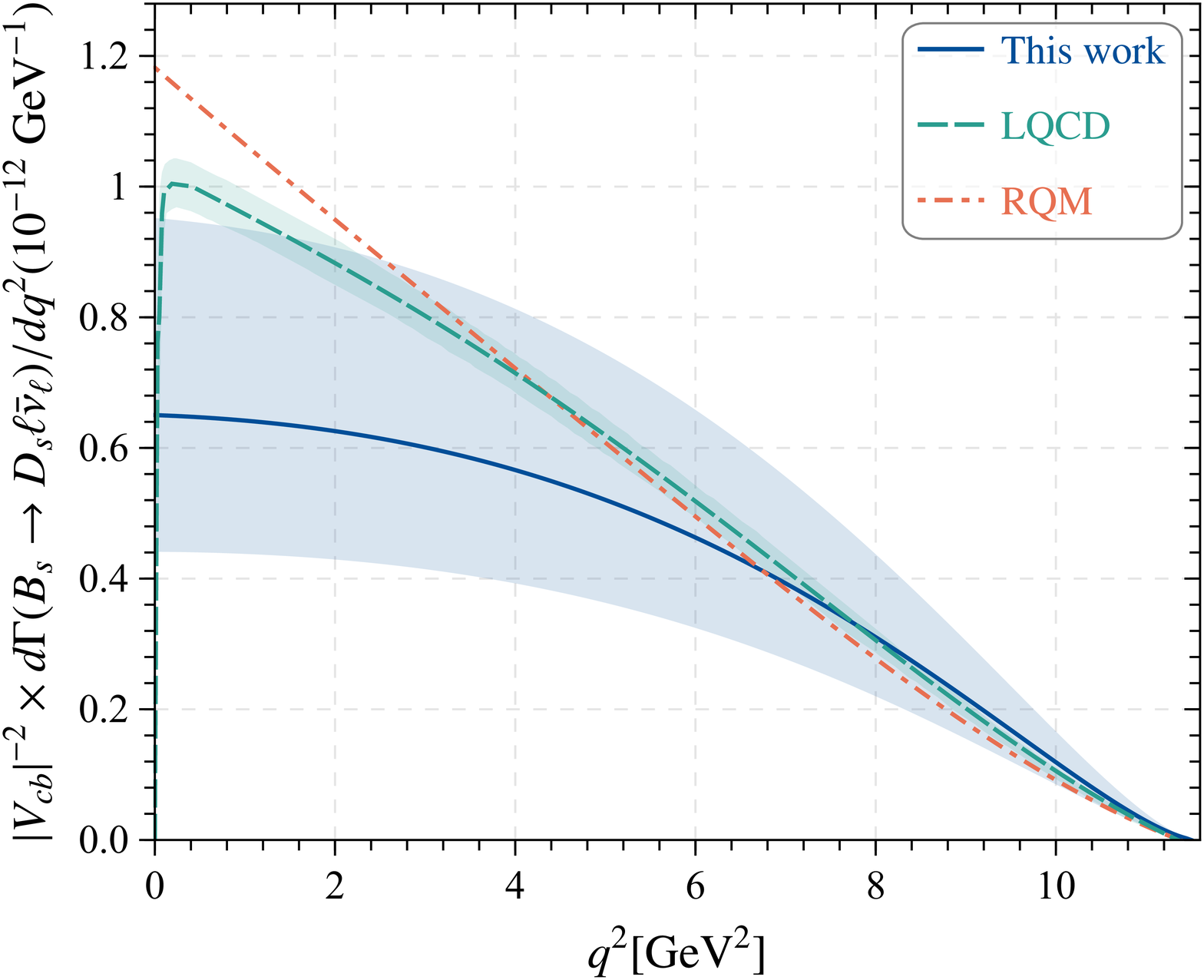}
\includegraphics[width=0.45\textwidth]{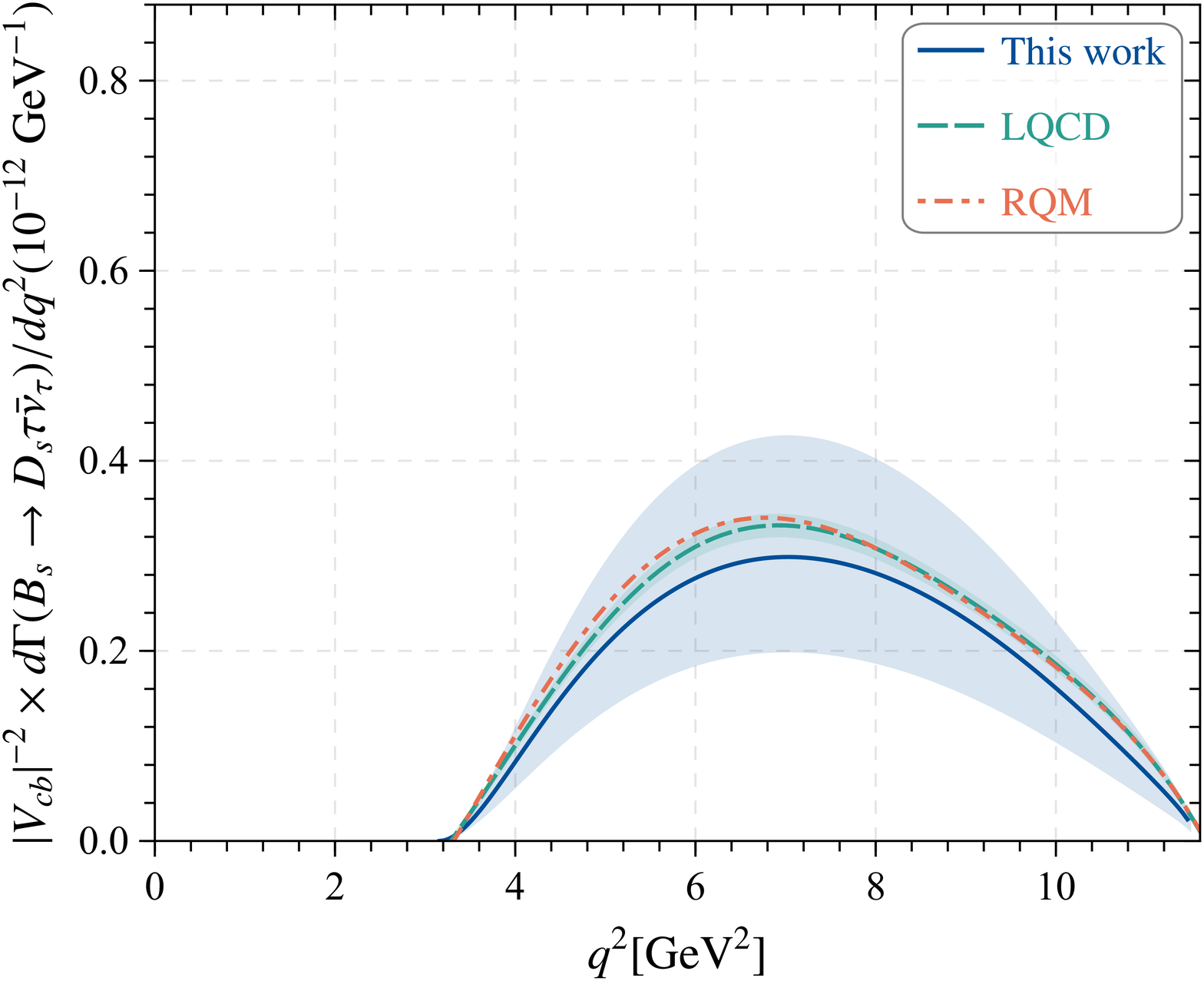}
\end{center}
\caption{The differential decay widths of $B_s \to D_s \ell \bar{\nu}_\ell$, where the uncertainties are squared averages of those from all the mentioned error sources. The predictions under the LQCD approach~\cite{McLean:2019qcx} and the RQM approach~\cite{Faustov:2012mt} are also presented.}
\label{width1}
\end{figure}

We present the differential decay widths of $B_s \to D_s \tau\bar \nu_\tau$ and $B_s \to D_s \ell^\prime\bar \nu_{\ell^\prime}$ in Figure~\ref{width1}, in which the solid lines are for the central choices of input parameters, and the shaded bands are uncertainties by adding all the errors caused by the error sources such as $f^{B_s\to D_s}_{+,0}(q^2)$, $m_{B_s}$, $m_{D_s}$, $|V_{cb}|$, $G_F$ and $m_\ell$, etc., in quadrature. In addition, the predictions under the RQM approach~\cite{Faustov:2012mt} and the LQCD approach~\cite{McLean:2019qcx} are also given. One may observe that our prediction of $d\Gamma(B_s \to D_s \tau\bar \nu_\tau )/dq^2$ is consistent with the LQCD and RQM predictions in Refs.~\cite{McLean:2019qcx, Faustov:2012mt}; And for $d\Gamma(B_s \to D_s \ell^\prime\bar \nu_{\ell^\prime} )/dq^2$, our prediction agrees with the LQCD and RQM predictions~\cite{McLean:2019qcx, Faustov:2012mt} in larger $q^2$ region, but is smaller than those predictions in lower $q^2$ region.

\begin{table}[htb]
\caption{Theoretical predictions of the branching fractions $\mathcal{B}(B_s\to D_s \ell^{\prime}\bar\nu_{\ell^{\prime}})$ and $\mathcal{B}(B_s\to D_s\tau\bar\nu_\tau)$ (in unit: $10^{-2}$).}
\begin{tabular}{l l l}
\hline
~Methods~~~~~~~~~ & ~$\mathcal{B}(B_s\to D_s \ell^{\prime} \bar{\nu}_{\ell^{\prime}})$~ & ~$\mathcal{B}(B_s\to D_s \tau \bar{\nu}_\tau)$~ \\
\hline
~This work~(HQEFT)~ & ~$1.817^{+0.802}_{-0.571}$~ & ~$0.606^{+0.266}_{-0.211}$~ \\
~pQCD~\cite{Hu:2019bdf}~ & ~$1.97^{+0.89}_{-0.51}$~ & ~$0.72^{+0.32}_{-0.23}$~ \\
~pQCD+LQCD~\cite{Hu:2019bdf}~ & ~$1.84^{+0.77}_{-0.51}$~ & ~$0.63^{+0.17}_{-0.13}$~ \\
~pQCD~\cite{Fan:2013kqa}~ & ~$2.13^{+1.12}_{-0.77}$~ & ~$0.84^{+0.38}_{-0.28}$~ \\
~RQM~\cite{Faustov:2012mt}~ & ~$2.1 \pm 0.2$~ & ~$0.62 \pm 0.05$~ \\
~RQM~\cite{Bhol:2014jta}~ & ~$2.54^{+0.28}_{-0.27}$~ & ~$0.695^{+0.085}_{-0.075}$~ \\
~CQM~\cite{Zhao:2006at}~ & ~$2.73-3.00$~ & ~$-$~ \\
~QCDSR~\cite{Blasi:1993fi}~ & ~$2.46 \pm 0.38$~ & ~$-$~ \\
~QCDSR~\cite{Azizi:2008tt}~ & ~$2.8-3.5$~ & ~$-$~ \\
~LCSR~\cite{Zhang:2021wnv}~ & ~$2.03^{+0.35}_{-0.49}$~ & ~$-$~ \\
~LCSR~\cite{Li:2009wq}~ & ~$1.0^{+0.4}_{-0.3}$~ & ~$0.33^{+0.14}_{-0.11}$~ \\
~LQCD~\cite{Dutta:2018jxz}~ & ~$2.013-2.469$~ & ~$0.619-0.724$~ \\
~BSE~\cite{Chen:2011ut}~ & ~$1.4-1.7$~ & ~$0.47-0.55$~ \\
\hline
\end{tabular}
\label{branch}
\end{table}

We present the branching fractions $\mathcal{B}(B_s\to D_s \ell^{\prime}\bar\nu_{\ell^{\prime}})$ and $\mathcal{B}(B_s\to D_s\tau\bar\nu_\tau)$ in Table~\ref{branch}, where the predictions under various approaches are also presented as a comparison. It is noted that our present predictions are consistent with most of the previous predictions within errors. Especially, our prediction of $\mathcal{B}(B_s\to D_s \ell^{\prime}\bar\nu_{\ell^{\prime}})$ is in good agreement with the pQCD prediction of Refs.\cite{Fan:2013kqa, Hu:2019bdf} and the pQCD+LQCD approach~\cite{Hu:2019bdf}, and our prediction of $\mathcal{B}(B_s\to D_s\tau\bar\nu_\tau)$ is in good agreement with the pQCD+LQCD prediction~\cite{Hu:2019bdf} and the RQM predictions of Refs.\cite{Faustov:2012mt, Bhol:2014jta}.

\begin{table}[htb]
\caption{The ratios $\mathcal{R}(D_s)$ under various approaches. }
\begin{tabular}{l l}
\hline
~Methods~~~~~~~~~~~~~~~~~~~~~~~~~~~~~~~~~~~~~~~~~~~~~ & ~$\mathcal{R}(D_s)$~ \\
\hline
~This work~(HQEFT)~ & ~$0.334\pm 0.017$~ \\
~pQCD~\cite{Hu:2019bdf}~ & ~$0.365^{+0.009}_{-0.012}$~ \\
~pQCD+LQCD~\cite{Hu:2019bdf}~ & ~$0.341^{+0.024}_{-0.025}$~ \\
~pQCD~\cite{Fan:2013kqa}~ & ~$0.392\pm0.022$~ \\
~RQM~\cite{Faustov:2012mt}~ & ~$0.295$~ \\
~RQM~\cite{Bhol:2014jta}~ & ~$0.274^{+0.020}_{-0.019}$~ \\
~LQCD~\cite{Dutta:2018jxz}~ & ~$0.299^{+0.027}_{-0.022}$~ \\
~LQCD~\cite{Monahan:2017uby}~ & ~$0.314\pm0.006$~ \\
~CCQM~\cite{Soni:2021fky}~ & ~$0.271 \pm 0.069$~ \\
~LCSR~\cite{Li:2009wq}~ & ~$0.33$~ \\
\hline
\end{tabular}
\label{ratio}
\end{table}

Combining Eqs.~\eqref{Eq:BF}, \eqref{Eq:DDF1} and \eqref{Eq:DDF2}, we can obtain the ratio $\mathcal{R}(D_s)$
\begin{eqnarray}\label{Rpi}
\mathcal{R}(D_s) = \frac{\int^{q^2_{max}}_{m^2_{\tau}}d\Gamma(B_s\to D_s \tau \bar\nu_\tau)/ dq^2}{\int^{q^2_{max}}_{0}d\Gamma (B_s\to D_s l^{\prime} \bar\nu_{l^{\prime}})/ dq^2},
\end{eqnarray}
which leads to
\begin{eqnarray}\label{Rp}
\mathcal{R}(D_s) =0.334 \pm 0.017.
\end{eqnarray}
We present the ratios under various approaches in Table~\ref{ratio}. And to be consistent with the above branching fractions, our ratio $\mathcal{R}(D_s)$ is in good agreement with prediction under the pQCD+LQCD approach~\cite{Hu:2019bdf}.

\section{summary}\label{Section:IV}

In the present paper, we make a detailed study on the TFFs of the semileptonic decay $B_s\to D_s \ell\nu_\ell$ under the LCSR approach within the framework of HQEFT. By using the chiral correlator, the TFFs $f_{+,0}^{B_s \to D_s}(q^2)$ are dominated by the leading-twist contributions and the accuracy of the LCSR prediction is improved. At the maximum recoil point, we have $f_{+,0}^{B_s \to D_s}(0)=0.533^{+0.112}_{-0.094}$. After applying the $z$-series extrapolation, we obtain the TFFs in the whole physical $q^2$-region. Figure~\ref{fp} and Figure~\ref{width1} show the extrapolated TFFs $f_{+,0}^{B_s \to D_s}(q^2)$ and the differential decay widths of $B_s \to D_s \ell \bar{\nu}_\ell$, respectively. Furthermore, we derive the branching fractions $\mathcal{B}(B_s \to D_s \ell \bar\nu_\ell)=(1.817^{+0.802}_{-0.571}) \times 10^{-2}$ and $\mathcal{B}(B_s \to D_s \tau \bar\nu_\tau) = (6.061^{+2.660}_{-2.114})\times 10^{-3}$. The resultant ratio $\mathcal{R}(D_s)=0.334\pm 0.017$ agrees well with the previous prediction under a combined approach of pQCD+LQCD~\cite{Hu:2019bdf}. This could be treated as a good example of showing the consistency of the TFFs under various approaches~\cite{Huang:2004hw}. Analyzing the data in Tab.~\ref{ratio}, we can find that the predictions of $\mathcal{R}(D_s)$ through various methods are not in good agreement with each other, which needs more reasonable and accurate research in the future. At the same time, we also look forward to the experimental measurements of $\mathcal{R}(D_s)$, so as to test the theoretical prediction for $\mathcal{R}(D_s)$ in the framework of SM.

\hspace{1cm}

{\bf Acknowledgments}: We are grateful to Rui-Yu Zhou for helpful discussions. This work was supported in part by the National Science Foundation of China under Grant No.11875122, No.11947406, No. 12175025 and No. 12147102, the Project of Guizhou Provincial Department of Science and Technology under Grant No.KY[2019]1171 and No.ZK[2021]024, the Project of Guizhou Provincial Department of Education under Grant No.KY[2021]030 and No.KY[2021]003, and by the Chongqing Graduate Research and Innovation Foundation under Grant No.ydstd1912.

\end{document}